\documentclass[11pt,a4paper]{article}
\usepackage{graphicx}
\usepackage{multirow}
\usepackage{bm}
\usepackage{amsmath}
\DeclareMathOperator*{\argmax}{arg\,max}
\usepackage[hyperref]{acl2021}
\usepackage{times}
\usepackage{latexsym}

\usepackage{microtype}

\aclfinalcopy 

\title{A Semi-supervised Multi-task Learning Approach to Classify Customer Contact Intents}

\author{Li Dong \\
  Amazon \\
  \texttt{ldonga@amazon.com} \\\And
  Matthew C. Spencer \\
  Amazon \\
  \texttt{matsp@amazon.com} \\\And
  Amir Biagi \\
  Amazon \\
  \texttt{biagiab@amazon.com}\\}

\date{}

\begin{document}
\maketitle
\begin{abstract}
In the area of customer support, understanding customers' intents is a crucial step. Machine learning plays a vital role in this type of intent classification. In reality, it is typical to collect confirmation from customer support representatives (CSRs) regarding the intent prediction, though it can unnecessarily incur prohibitive cost to ask CSRs to assign existing or new intents to the  mis-classified cases. Apart from the confirmed cases with and without intent labels, there can be a number of cases with no human curation. This data composition (Positives + Unlabeled + multiclass Negatives) creates unique challenges for model development. In response to that, we propose a semi-supervised multi-task learning paradigm. In this manuscript, we share our experience in building text-based intent classification models for a customer support service on an E-commerce website. We improve the performance significantly by evolving the model from multiclass classification to semi-supervised multi-task learning by leveraging the negative cases, domain- and task-adaptively pretrained ALBERT on customer contact texts, and a number of un-curated data with no labels. In the evaluation, the final model boosts the average AUC ROC by almost 20 points compared to the baseline finetuned multiclass classification ALBERT model.

\end{abstract}

\section{Introduction}
As machine learning makes rapid advances in the area of natural language processing (NLP), it is becoming more common to aid customer support representatives (CSRs) with NLP models. This not only ensures timely and consistent replies to customers, but also reduces operational costs for organizations. We can see successful use cases from organizations such as Alibaba \cite{Fu2020}, Uber \cite{molino2018}, Square \cite{fotso2018}, AT\&T \cite{gupta2010}, IBM \cite{mani2018}, Los Alamos National Laboratory \cite{deLucia2020}, and US Navy \cite{powell2020}. In general, identifying the intents of the coming contacts is the first step in customer support. Therefore, accurate intent classification is crucial.

\begin{figure}[h]
	\centering
	\includegraphics[width=8cm]{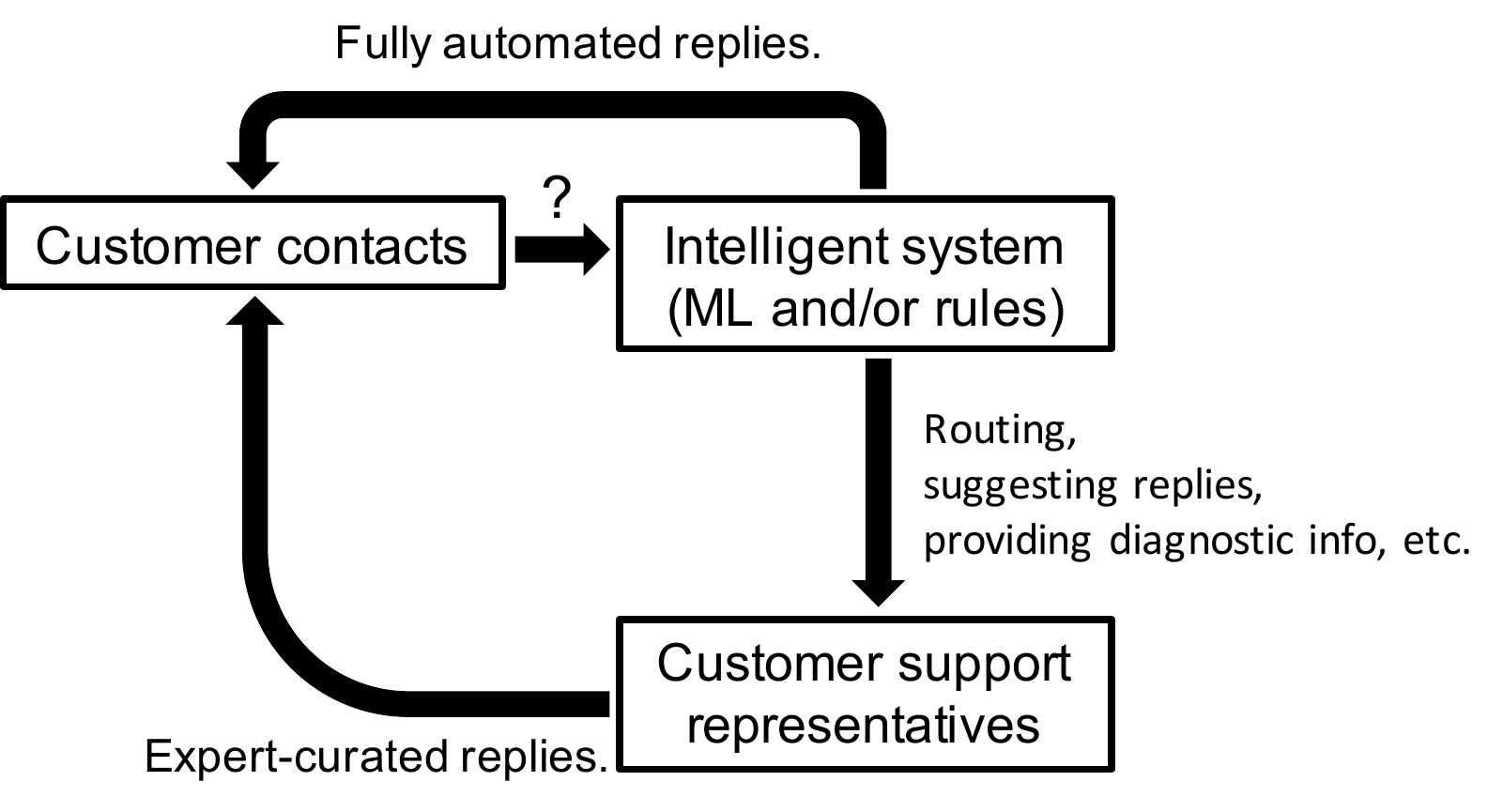}
	\caption{Intelligent customer support loop}
	\label{fig:customer support}
\end{figure}

Intent classification is a broad topic mostly falling under the umbrella of NLP. In this manuscript, we limit our discussion to intent classification in the area of customer support. In the past two decades, researchers have been trying to improve the efficiency of customer support by detecting customer intents with machine learning approaches \cite{molino2018, powell2020, deLucia2020, hui2000, gupta2010, fotso2018, mani2018, Sarikaya2011, Gupta2006, Xu2013}. We can loosely categorize these approaches into text classification \cite{molino2018, powell2020, deLucia2020, hui2000, gupta2010, fotso2018}, question-answer (QA) system \cite{mani2018} and automatic speech recognition (ASR) \cite{Sarikaya2011, Gupta2006, Xu2013}. In this manuscript, we focus on using text classification methods to classify intents for customer support. To deal with unstructured text data, researchers use handcrafted features \cite{hui2000, gupta2010}, Bag-of-Words type of features \cite{powell2020}, features from topic modeling \cite{deLucia2020} and vectorization type of features, such as word2vec \cite{fotso2018, molino2018} and doc2vec \cite{deLucia2020}. By consuming these features, classifiers determine the intent of a case and the case can be routed to specialists \cite{molino2018, gupta2010, deLucia2020, powell2020} and/or a reply template from the ``Answer Bank'' can be provided \cite{molino2018, fotso2018, hui2000}. A general intelligent customer support loop can be seen in Figure \ref{fig:customer support}.

To meet ever-changing business needs, the intent taxonomy is generally under active development \cite{molino2018, fotso2018, deLucia2020}. It is not realistic to manually relabel all cases after each intent taxonomy update. This means that we have positive cases (P) with assigned intents and unlabeled cases (U) in data. Moreover, to maintain a high standard of customer satisfaction, intent classification is typically a human-in-the-loop process \cite{Fu2020, molino2018, fotso2018, gupta2010, powell2020}. Specifically, the CSRs are asked to confirm the intent predictions, a process we refer to as ``curation'' in this manuscript. The negative cases (N) identified by CSRs are indeed hard cases, since their prediction scores are above the preset confidence threshold yet they are mis-classified by the existing model. It is an active research area to create classifiers with only P and U \cite{elkan2008, xu2017ijcai}. Some research has explored models that also include N, but they have been only concerned with binary classifiers \cite{fei2015social, hsieh2019, li2010negative}. 

In this manuscript, we adopt the semi-supervised paradigm and the multi-task approach to deal with the U and the multiclass N, respectively. Moreover, in contrast to the above-mentioned works about intent classification for customer support, we use the ALBERT pretrained language model \cite{lan2019} plus domain- and task-adaptive pretraining \cite{ramponi2020, gururangan2020} to process texts. In the following sections, we describe how these techniques improve the model performance. 

The paper outline is as follows. We start with Section \ref{background} by elaborating the business background and how we pose it as a machine learning problem. Then we describe the details of the training data and models in Section \ref{method}, compare the models by conducting experiments with real data in Section \ref{experiments and results}, and discuss the results in Section \ref{discussion}. We conclude in Section \ref{conclusion}.

\section{Background} \label{background}

The E-commerce website of interest receives many support requests from customers in each second. There is a team of CSRs to actively address the requests via phone, online chat, and email channels. Identifying appropriate requests and grouping them into categories is not a trivial task. While a deep discussion of the taxonomy building process is out of the scope of this manuscript, it is sufficient to know that we have a taxonomy system that is similar to those described in \cite{molino2018, fotso2018}, where customized reply templates are pre-compiled for each customer contact intent. This study elaborates our journey building machine learning models to classify the intents.

\section{Methodology} \label{method}

Since the inception of BERT \cite{devlin2019}, an abundance of research in the area of NLP has demonstrated it to be an effective approach to transfer knowledge from pretrained language models to downstream tasks \cite{xia2020, wang2018, wang2019, rajpurkar2016, lai2017race}. Following BERT's architecture, there is a stream of research that achieve comparable or better performance, to name a few \cite{lan2019, liu2019, wang2019structbert, Clark2020, yang2019xlnet, Sanh2019}. Among these BERT variants, ALBERT aims to strike a balance between model performance and model size \cite{lan2019}. Therefore, we use albert-base-v2 as the backbone encoder and perform further pretraining and finetuning. The implementation is based on Transformers from Huggingface \cite{Wolf2019}.

\subsection{Training Data}

\subsubsection{Features}

The input to the model is a collection of emailed support requests in text format. The texts are minimally preprocessed, including removing invalid characters, lowercasing letters and replacing some obvious entities with consistent words, such as replacing urls and emails to url\_id and email\_id.

\subsubsection{Targets} \label{targets}

In industrial machine learning applications, it is typical to construct a feedback loop to collect training data. In most cases, it is straightforward to obtain a simple ``yes'' or ``no'' from human labelers with respect to the predictions. That means the human labelers only need to accept or dismiss the recommendations. Those ``yes'' cases are confirmed positive ones with explicit labels. However, in a common multiclass classification setting, the ``no'' cases can have any other label, so the labels are effectively unknown. In some scenarios, such as object detection in computer vision, it is not hard to ask human labelers to assign a label to those negative cases. However, it is highly non-trivial to ask for a valid label in many NLP applications, owing to the size of the taxonomy and the necessary domain-expertise, as is the case for the intent classification in this manuscript. Therefore, in our training data, we only have ``yes'' or ``no'' feedback to each case in each intent class. 

Since the scope of the intent taxonomy is not trying to cover all customer support requests, there are many requests falling out of the scope of the taxonomy but still scored by the model. The negative cases are either out-of-scope requests or in-scope requests falling in the wrong bucket. The former one is more probable, since the requests are false positives for existing classes with high confidence scores above the preset thresholds. In this manuscript, we tried two ways to deal with this situation. 

\begin{enumerate}
	\item We can simply exclude the negative cases from training data, since they do not come with labels. In this scenario, it is a multiclass classification model trained on positive cases, i.e. confirmed intents. However, we lose valuable signals by excluding the negative cases.
	
	\item Since the negative cases are indeed hard negatives and contain valuable signals, we can use the multi-task learning paradigm to elegantly treat the negatives for each intent class as the negative samples for a binary classification task. In this scenario, we have a binary classification task for each class plus a multiclass classification task for all classes. It is also not necessary to examine the negative cases and assign them to appropriate new or existing classes, especially when the labeling efforts outweigh the benefits it could bring to model development. With this approach, we make full use of the signals in the training data.
	
\end{enumerate}

Apart from the multiclass positive (P) and negative (N)  cases mentioned above, we also have the un-curated cases that do not come with labels, i.e. the U cases. We adopt an iterative semi-supervised approach to deal with them. The approach is described in Section \ref{ss mt dtapt albert}.

\subsection{Models} \label{model evolution}

\subsubsection{ALBERT} \label{albert}

Following the pretraining-finetuning framework for language models, we start with a finetuned ALBERT. We simply remove the masked language model (MLM) head and the sentence order prediction (SOP) head from ALBERT and add a sequence classification head. Following the convention from \cite{devlin2019}, the final hidden vector corresponding to the first input token [CLS] is used for classification. We denote this vector as the classification vector in the rest of the manuscript. We note that this ALBERT model is trained as a multiclass classification with only \textit{positive} cases.

\subsubsection{SS MT D/TAPT ALBERT} \label{ss mt dtapt albert}

The pretrained language models are mostly trained on well-known corpora, such as Wikipedia, Common Crawl, BookCorpus, Reddit, etc. However, in many cases, we need to apply the language models to very different domains, like BioMed, scientific publication, or product reviews. For these types of problems, researchers have found that, in addition to finetuning on specific downstream tasks, it is beneficial to adapt the language models to the domain- and task-specific corpus, i.e. domain-adaptive pretraining (DAPT) and task-adaptive pretraining (TAPT) \cite{gururangan2020}. This is achieved by further training the language modeling tasks, such as MLM, with the corpus of the domain and the task. We note that it can be difficult to rigorously define \textit{domain} in NLP. For the DAPT training in this manuscript, we simply use customer contacts in the past few months as the domain corpus and follow the training recommendations from \cite{gururangan2020}.

\begin{figure}[h]
	\centering
	\includegraphics[width=\columnwidth]{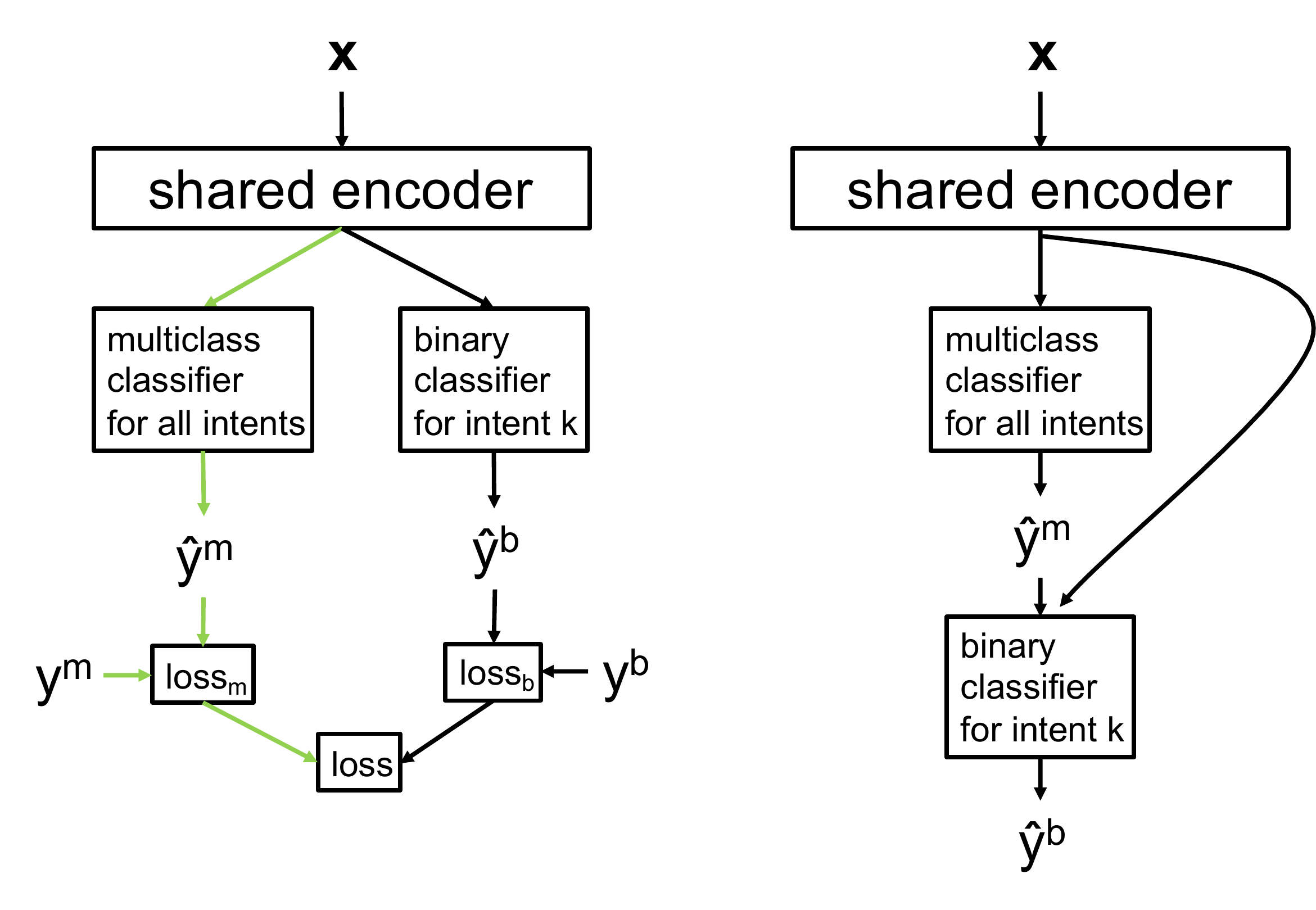}
	\caption{Training (left) and inferencing (right) for the multi-task learning strategy, where $k \in [0 ... n - 1]$,  and $n$ is the number of intent classes. In training time, the green path is only executed when $\textbf{x}$ is a positive sample.}
	\label{fig:albert_arch}
\end{figure}

To make full use of the feedback from CSRs, we include the negatively confirmed cases and treat each class as a separate binary classification task in addition to the multiclass classification task. We accomplish the modeling with the multi-task (MT) learning paradigm \cite{liu2019}. In this case, we have $n + 1$ tasks, i.e. $n$ binary classification tasks and $1$ multiclass classification task. As illustrated on the left of Figure \ref{fig:albert_arch}, we train the model in an end-to-end fashion. This means the $n + 1$ tasks are finetuned jointly sharing the same encoder.  We note that every positive sample belongs to two tasks (the multiclass classification task and one binary task) and each negative sample only belongs to the corresponding binary classification task. In inferencing time, as illustrated on the right of Figure \ref{fig:albert_arch}, the model first processes the case text through the encoder to get the classification vector. Then the multiclass classification task consumes the vector and predicts the class. In the end, the same vector is routed to the binary task corresponding to that class, predicting the probability of the intent class accepted by the CSRs. 

To make it more concrete, we can see the training loss implementation in Equation (\ref{eq:trn}). $y^b$  is the binary label, i.e. $1$ means it is a positive sample and its intent class is confirmed by CSRs with ``yes''. $l^m$ is the multiclass task loss. $\bm{y}^m$ is the one-hot encoded n-dimensional multiclass label vector. $\bm{l}^b$ is the loss function vector for $n$ binary tasks. $N$ is the number of samples. Typical cross-entropy loss is used for all tasks here.

\begin{equation} \label{eq:trn}
\mathcal{L} = \frac{1}{N} \sum_{i=1}^N (y_i^b \cdot l_i^m + \bm{y}_i^m \cdot  \bm{l}_i^b)
\end{equation}

For the inferencing process, we refer to Equations (\ref{eq:inf1})-(\ref{eq:inf4}). $\bm{x}$ is the tokenized sequence vector. $\bm{u}$ is the classification vector, i.e. the embedding vector for the CLS token. $f^m$ is the multiclass classifier. $f^b_k$ is the binary classifier for intent class $k$.

\begin{align}
	\bm{u} &= \textit{Encoder}(\bm{x}) \label{eq:inf1} \\
	\bm{\hat{y}}^m &= f^m(\bm{u}) \\
	k &= \argmax_i \bm{\hat{y}}^m(i),  i \in [0...n-1] \\
	\hat{y}^b &= f^b_k(\bm{u}) \label{eq:inf4}
\end{align}

Moreover, we add the semi-supervised (SS) strategy to take advantage of the un-curated data. While a large volume of model predictions are reviewed by the CSRs each second, we believe there are still a number of qualified cases that we miss. Therefore, we can train the model, make prediction on the un-curated cases, choose the high-confidence ones, and re-train the model with the labeled data plus the high-confidence cases. We follow this in an iterative manner until the improvement diminishes such that it cannot justify the training cost. We note that we only augment the data of the multiclass classification task and the data for the binary classification tasks remain unchanged throughout the iterative process. The same strategy is recently used by \cite{Schick2020} to create small language models that have similar performance to BERT and \cite{xie2020} to achieve state-of-the-art performance on Imagenet in computer vision.

Adding up the techniques described above, we denote this model as SS MT D/TAPT ALBERT.

\section{Experiments and Results} \label{experiments and results}

\subsection{Data and Experimental Setup}

\begin{table*} 
	\caption{Sample training data and how different training strategies incorporate them.}
	\label{tab:sample data}
	\begin{center}
		\begin{scriptsize}
			\begin{tabular}{l | l | l | l l | l l l l l}
				\hline
				\multirow{2}{*}{Curation} & \multirow{2}{*}{Composition} & \multicolumn{1}{c|}{Features} & \multicolumn{2}{c|}{Targets} & \multicolumn{5}{|c}{Training data for} \\
				& & Messages & Intents & CSR responses & Multiclass task & Binary tasks & DAPT & TAPT & SS \\
				\hline
				\multirow{4}{*}{Curated} & \multirow{2}{*}{Positives} & Could you help me? & General inquiry & Yes & Yes & Yes (+) & Yes & Yes & Yes \\
				& & How to setup account? & Account issue & Yes & Yes & Yes (+) & Yes & Yes & Yes \\
				\cline{2-10}
				& \multirow{2}{*}{Negatives} & How much is this? & Account issue & No & No & Yes (-) & Yes & Yes & Yes \\
			    & & Can you fix this issue? & General inquiry & No & No & Yes (-) & Yes & Yes & Yes \\
			    \hline
			    \multirow{2}{*}{Un-curated} & \multirow{2}{*}{Unlabeled} &  What's this? & General inquiry & N/A & No & No & Yes & No & Yes \\
			    & & Please help. & N/A & N/A & No & No & Yes & No & Yes \\
				\hline
			\end{tabular}
		\end{scriptsize}
	\end{center}
\end{table*}

For confidentiality reasons, we can only share directional numbers about the training data.  In this study, we consider 9 customer intent classes. The curated data is unbalanced among classes, ranging from a few thousand to tens of thousands of records per class. The class with the most samples is roughly 40 times as much as the class with the least samples. For each class, the ratio of positive-to-negative cases in the curated data is about 4. The un-curated data is roughly 20 times of the curated data. We use both the curated and un-curated data for DAPT and only curated data for TAPT. In the semi-supervision process, for each class, we select high-confidence samples from the un-curated data in each iteration to be roughly two to three times of the volume of the labeled samples in the curated data. Table \ref{tab:sample data} shows a few sample training data with dummy features and intents. The last column shows how different strategies incorporate them in training. 

After being processed with the ALBERT tokenizer, the total data amounts to about 800 million tokens with an average of about 80 per sample. We performed all experiments on Sagemaker on AWS. We used 2 ml.p3.16xlarge instances with distributed data parallelism for DAPT and TAPT, 1 ml.p3.8xlarge instance for finetuning language models, and 1 ml.p3.8xlarge instance for batch inferencing testing data.

We hold out a portion of the data as development data to tune hyperparameters. We follow the suggestions from \cite{gururangan2020, liu2019} for DAPT and TAPT and \cite{devlin2019} for finetuning. For the end-to-end multi-task learning process, we kept a unit weight for each task and did not explore different weight combinations. More research about tuning task weights in multi-task learning can be found in \cite{Cipolla2018}.

\subsection{Evaluation}

\subsubsection{Pretrained models} \label{pretrained models}

In this section, we evaluate the performance of the pretrained language models, the out-of-the-box ALBERT and the D/TAPT ALBERT. We note that the pretrained language models are evaluated before any finetuning happens. 

\begin{figure}[h]
	\centering
	\includegraphics[width=\columnwidth]{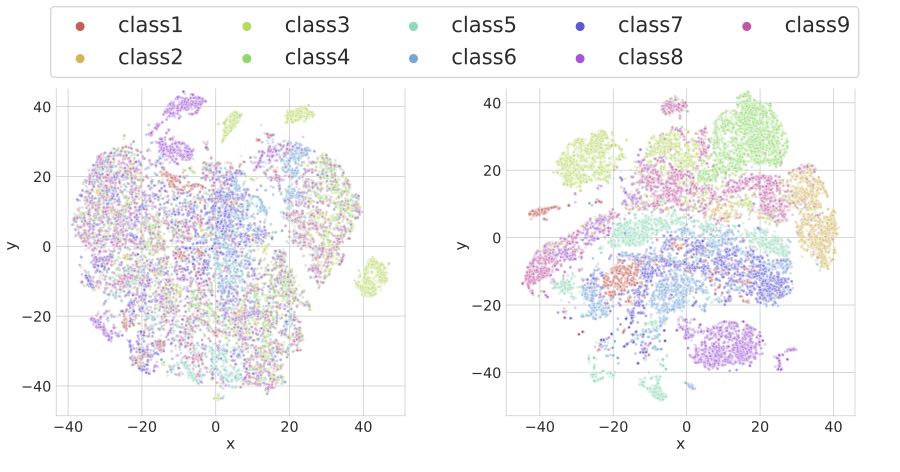}
	\caption{t-SNE plots using the dimension-reduced classification vectors from the off-the-shelf pretrained ALBERT (left) and the D/TAPT ALBERT (right).}
	\label{fig:tsne}
\end{figure}

To visually demonstrate how the adaptive pretraining improves the clustering performance of the classification vector, we sample a couple thousand cases per class and apply t-SNE \cite{VanDerMaaten2008} to the reduced classification vector for each case. We reduce the dimension of the classification vectors from 768 to 50 with PCA to keep the computational cost of t-SNE in check. In Figure \ref{fig:tsne}, we can see how clustering improves from the vanilla ALBERT on the left to D/TAPT ALBERT on the right.

To more quantitatively assess the performance of the off-the-shelf pretrained ALBERT and the D/TAPT ALBERT, we sample a couple thousand cases per class and use k-nearest-neighbor classifiers (kNN) to predict each sample's class based on its $k$ neighbors. We use the Euclidean distance between the classification vector for each case as the similarity metric for kNN. We compute the average accuracy by varying $k$ from 3 to 99 in interval of 2 and report it in Table \ref{tab:knn}. As a result, D/TAPT lifts the accuracy by more than 30 points compared to the vanilla ALBERT. Similar performance lift is also observed in \cite{reimers2019}. This illustrates that D/TAPT can improve the clustering performance of the classification vector when the clustering rules are closely related to the domain corpus. The absolute accuracy values are not reported here for confidentiality reasons.

\begin{table} 
	\caption{Average kNN prediction accuracy using the classification vectors from the pretrained models}
	\label{tab:knn}
	\begin{center}
		\begin{small}
			\begin{tabular}{ccc}
				\hline
				ALBERT & D/TAPT ALBERT \\
				\hline
				0\% & +33\% \\
				\hline
			\end{tabular}
		\end{small}
	\end{center}
\end{table}

\subsubsection{Finetuned models} \label{finetuned models}

In practice, for each class, we expect to route more positive cases and less negative cases to our CSRs with machine learning models. That means we expect our models to better differentiate positives from negatives for each class. Area Under the Curve - Receiver Operating Characteristics (AUC ROC) is a natural metric for such binary classification problem. We note that the commonly-used \textit{accuracy} metric is not appropriate in this context since the negatives do not have ground truth labels in our data. The evaluation data is from recent few weeks. For confidentiality reasons, we hide the axis for AUC ROC and make the values relative to the baseline finetuned ALBERT model for each class.

\begin{figure}[h]
	\centering
	\includegraphics[width=\columnwidth]{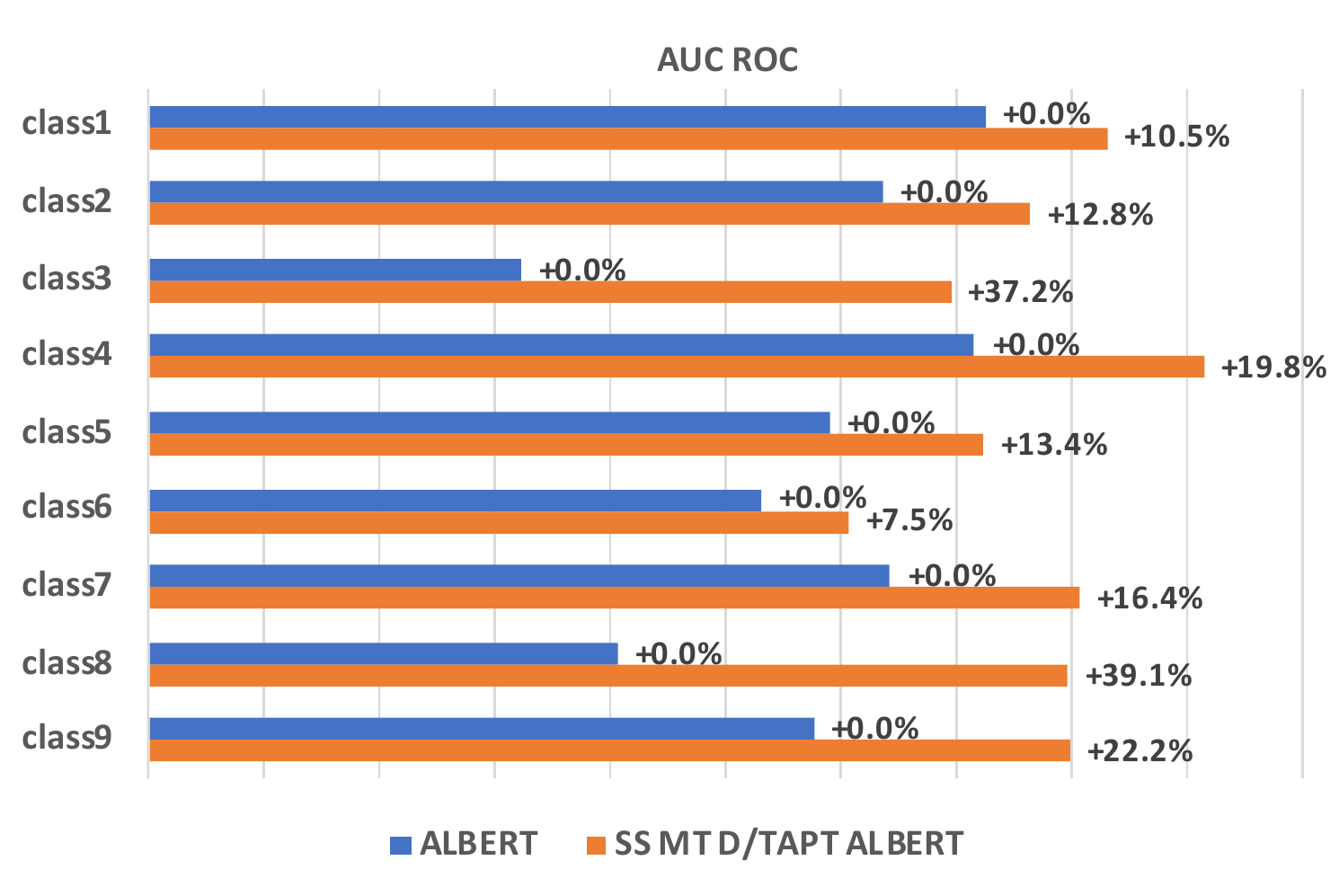}
	\caption{The AUC ROC of each class for finetuned ALBERT and SS MT D/TAPT ALBERT}
	\label{fig: auc roc labels}
\end{figure}

\begin{table} 
	\caption{The average and sample-weighted average AUC ROC for different experiment settings}
	\label{tab:auc roc all}
	\begin{center}
		\begin{small}
			\begin{tabular}{lcc}
				\hline
				Model & avg AUC ROC & wavg AUC ROC \\
				\hline
				ALBERT & +0\% & +0\% \\
				+ MT & +17.8\% & +14.3\% \\
				+ MT DAPT & +18.4\% & + 15.8\% \\
				+ MT D/TAPT & +19.0\% & +16.1\% \\
				+ SS MT D/TAPT & +19.9\% & +17.0\% \\
				\hline
			\end{tabular}
		\end{small}
	\end{center}
\end{table}

In Figure \ref{fig: auc roc labels}, for each class, we can observe consistent improvement of SS MT D/TAPT ALBERT over finetuned ALBERT in terms of AUC ROC. Overall, the SS MT D/TAPT ALBERT model brings 19.9 points increase in average AUC ROC and 17 points increase in sample-weighted average AUC ROC, compared to the finetuned ALBERT model.

Furthermore, it is interesting to see how each strategy in the SS MT D/TAPT ALBERT model contributes to the performance improvement. In Table \ref{tab:auc roc all}, we show the average and sample-weighted average AUC ROC improvement by incrementally adding one strategy at a time. We can see that the MT strategy boosts the average AUC ROC by 17.8 points and the sample-weighted average AUC ROC by 14.3 points, compared to the finetuned ALBERT. This demonstrates the effectiveness of including negative signals with MT strategy. On top of MT, we apply DAPT, D/TAPT and SS incrementally. Each strategy pushes up the average and sample-weighted average AUC ROC by roughly 1 point.

\section{Discussion} \label{discussion}

Apart from processing the dismissed recommendations with this multi-task setting, there is another heuristic approach that is commonly adopted under this circumstance. We can group all the dismissed recommendations into an extra bucket \textit{Others} \cite{fotso2018}. The advantage of this approach is that we can pose the problem as a straightforward multiclass classification. The disadvantage is that the dismissed recommendations can either be mis-classified and belong to other existing classes, or belong to unknown classes that might be included in the future taxonomy. In the former scenario, the dismissed recommendations create noise for their true class and the \textit{Others} class; In the latter scenario, the dismissed recommendations can seemingly improve performance for current taxonomy, while they can pollute the future training when the unknown classes are launched in the updated taxonomy. In both scenarios, grouping the dismissed recommendations into \textit{Others} can negatively impact the training. 

In terms of computational cost, both adaptive pretraining and semi-supervision consume a considerable amount of power, since the former is typically trained on the MLM task through a large corpus and the latter is a iterative finetuning and inferencing process where the data for inferencing are often in large volume. In the meantime, the MT strategy is a cost-effective way to improve model performance by considering negative samples. By examining Table \ref{tab:auc roc all}, compared to the baseline finetuned ALBERT, we can see the MT strategy increases the average AUC ROC by 17.8 points while D/TAPT and SS add 2.1 points on top of that. The additional cost for the MT strategy, compared to the typical multiclass classification strategy, is simply a binary classifier for each class. It is negligible in both training and inferencing.

For the sake of easy implementation of the end-to-end multi-task training, we only feed the training data related to one task in each batch. In this way, we can keep the loss function for each task separate. It is possible that including data for various tasks in each batch can bring benefits to training. This assumption can be explored in future studies.

This study is only concerned with corpus in English. Similar modeling strategies can be followed for other high-resource languages which we have ample training data. However, as in the customer service departments of most global organizations, it is common to receive customer contacts in various low-resource languages, in which case the training data is scarce. Recent advances in cross-lingual language models, such as mBERT \cite{devlin2019}, XLM \cite{conneau2019}, Unicoder \cite{huang2019unicoder} and FILTER \cite{fang2020filter}, can shed light on this situation and we plan to investigate it in the future.

In the area of customer support, both \cite{molino2018} and \cite{fotso2018} propose neural networks that combine unstructured text features from customers' messages and structured features describing customers' interaction with the platforms. They empirically demonstrated benefits of including the latter feature group. The next step for our study is to evaluate the influence of the customer-website interaction features, when combining with advanced language models.

For the model candidates with multi-task strategy in this manuscript, we train all tasks jointly with an end-to-end multi-task deep learning approach, as described in the left plot of Figure \ref{fig:albert_arch}. We want to point out the isolating effect of the end-to-end training approach. In one experiment, we trained the tasks independently, i.e. we first trained the multiclass classification task with the off-the-shelf ALBERT, and then, for the binary tasks, we trained $n$ logistic regression binary classifiers with the classification vector from the multiclass classification task. We still achieved 12.2 and 8.2 points above the baseline in terms of average AUC ROC and sample-weighted average AUC ROC. On one hand, this shows that even training simpler models independently can still bring performance lifts, thus emphasizing the powerful signal brought by the negative cases; On the other hand, if compared to the ALBERT + MT model in Table \ref{tab:auc roc all}, it also shows the benefits of end-to-end training.

As in most machine learning applications, the actual model performance is determined by the choice of the operational point for each intent class and the operational point is determined from the precision-recall (PR) curve. For the sake of brevity, we ignore the PR plots because, for each class, the PR curve of the baseline ALBERT model is well under the envelop of the PR curve of the SS MT D/TAPT ALBERT model. This is expected due to the large boost presented in Figure \ref{fig: auc roc labels}. We note that the AUC ROC can be a decent indication of AUC PR when the data is not so skewed \cite{Davis2006}. Therefore, the SS MT D/TAPT ALBERT indeed outperforms the baseline for every choice of operational point.

\section{Conclusion} \label{conclusion}

In this manuscript, we demonstrated and discussed the model performance improvement brought by multi-task learning, adaptive pretraining for ALBERT, and semi-supervised learning in the application of customer support on an e-commerce website. We observe $\sim$20 points performance increase in average AUC ROC when comparing the final model to the baseline multiclass classification model. This paradigm can be particularly helpful when there is a feedback system collecting confirmation from labelers. Future studies can extend this paradigm to more complex situations, such as when the intent taxonomy is deeply hierarchical or considering more feedback information than simple ``yes" or ``no".

\section*{Acknowledgments}

The authors wish to thank Harsha Aduri and Jieyi Jiang for providing support in data preparation, Jasmine Qi and Ilknur Egilmez for providing comments for the manuscript. We also thank the anonymous reviewers for their valuable suggestions.

\bibliographystyle{acl_natbib}
\bibliography{custom}


\end{document}